\documentclass[aps,prl,preprint,superscriptaddress,longbibliography]{revtex4-1}
\usepackage{amssymb}
\usepackage[utf8]{inputenc}
\usepackage{graphicx}
\usepackage{bm,color,subfigure,amsmath,hyperref}

\begin{document}

\title{Symmetry-broken dissipative exchange flows in thin-film ferromagnets with in-plane anisotropy}

\author{Ezio~Iacocca}
\email{ezio.iacocca@colorado.edu}
\affiliation{Department of Applied Mathematics, University of Colorado, Boulder, Colorado 80309, USA}
\affiliation{Department of Physics, Division for Theoretical Physics, Chalmers University of Technology, 412 96 Gothenburg, Sweden}

\author{T.~J.~Silva}
\affiliation{National Institute of Standards and Technology, Boulder, Colorado 80305, USA}
\thanks{Contribution of the National Institute of Standards and Technology; not subject to copyright in the United States.}

\author{Mark~A.~Hoefer}
\affiliation{Department of Applied Mathematics, University of Colorado, Boulder, Colorado 80309, USA}

\begin{abstract}
Planar ferromagnetic channels have been shown to theoretically support a long-range ordered and coherently precessing state where the balance between local spin injection at one edge and damping along the channel establishes a dissipative exchange flow, sometimes referred to as a spin superfluid. However, realistic materials exhibit in-plane anisotropy, which breaks the axial symmetry assumed in current theoretical models. Here, we study dissipative exchange flows in a ferromagnet with in-plane anisotropy from a dispersive hydrodynamic perspective. Through the analysis of a boundary value problem for a damped sine-Gordon equation, dissipative exchange flows in a ferromagnetic channel can be excited above a spin current threshold that depends on material parameters and the length of the channel. Symmetry-broken dissipative exchange flows display harmonic overtones that redshift the fundamental precessional frequency and lead to a reduced spin pumping efficiency when compared to their symmetric counterpart. Micromagnetic simulations are used to verify that the analytical results are qualitatively accurate, even in the presence of nonlocal dipole fields. Simulations also confirm that dissipative exchange flows can be driven by spin transfer torque in a finite-sized region. These results delineate the important material parameters that must be optimized for the excitation of dissipative exchange flows in realistic systems.
\end{abstract}

\maketitle

\section{Introduction}

Spin current utilized as a source to excite magnetization dynamics has attracted significant research efforts in the past few years~\cite{Hoffmann2013}. In contrast to charge currents, spin currents describe the spatial flow of electron angular momentum in the form of quantum mechanical spin. Spin currents can be generated by a variety of means. For example, pure spin currents arise by charge-spin transduction in materials with strong spin-orbit coupling~\cite{Hoffmann2013,Manchon2015} as electrons of a given spin predominantly flow in a specific direction, leading to spin accumulation at the materials' edges. Utilizing this effect, current-induced magnetization dynamics have been demonstrated in devices based on a metallic / magnetic material bilayer~\cite{Liu2011,Liu2012,Demidov2012,Demidov2014,Jiang2015,Awad2016}. However, spin current transport in metals is limited by the spin diffusion length, typically on the order of hundreds of nanometers.

An alternative perspective is gained by recognizing that spin current is the Onsager reciprocal of spin precession~\cite{Tserkovnyak2002b}. Spin precession excited by means of spin currents has been experimentally demonstrated as the generation of small-amplitude spin waves in bilayers~\cite{Hoffmann2013,Demidov2012,Demidov2014,Awad2016}. Spin waves are typically defined as a perturbation about a uniform magnetization state whose coherence and energy are lost by scattering events that populate the dispersion relation and couple to lattice vibrations when in a thermal bath. This implies that the spin wave amplitude decays exponentially~\cite{Stancil2009} and, consequently, spin current transport in magnetic materials is limited by a spin wave propagation length that is inversely proportional to the magnetic damping.

\begin{figure*}[t]
\centering \includegraphics[trim={0in .25in 0in .25in}, clip, width=6.5in]{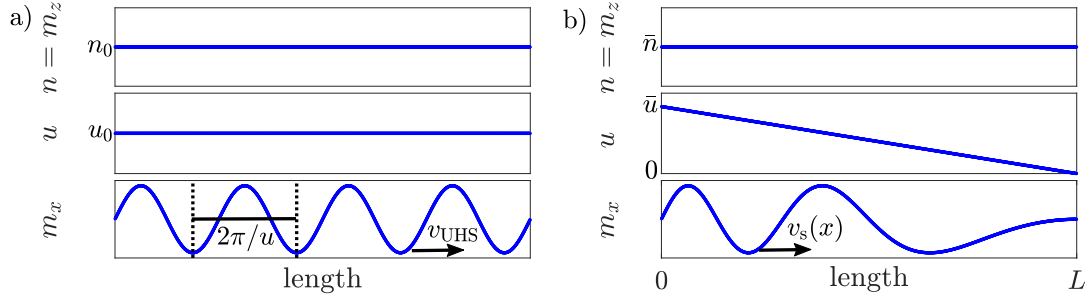}
\caption{ \label{figN} (color online) Schematic representation of the fluid density $n$, fluid velocity $u$, and magnetization component $m_x$. (a) A UHS parametrized by $-1<n_0<1$ and $0<u_0<1$, whose wavelength (vertical dashed lines) is set by the fluid velocity and $m_x$ translates with velocity $v_\mathrm{UHS}=-n_0/u_0$. (b) A dissipative exchange flow established in a planar ferromagnetic channel of length $L$ and spin injection $\bar{u}$ at the left edge exhibits a constant $\bar{n}<0$ and a linearly decaying fluid velocity. The fluid velocity's linear decay results in a space-dependent velocity $v_s(x)$ that increases towards the right edge and manifests as a space-dependent in-plane magnetization wavelength. }
\end{figure*}

Recent theoretical works have shown that magnetic materials support a fundamentally different magnetization state exhibiting a spatially homogenous precessional frequency that can pump dc spin currents into a suitable reservoir, such as an adjacent nonmagnetic metal. In their more general manifestation, planar magnetic materials in the conservative limit ($\alpha=0$) support extended uniform hydrodynamic states (UHSs)~\cite{Iacocca2017,Iacocca2017b} whereby the magnetization undergoes a spatial, large-amplitude rotation about the normal-to-plane axis. A notable feature of UHSs is that the magnetization is textured, i.e., non-collinear in neighboring sites, and establishes an equilibrium exchange flow~\cite{Bruno2005} that can be analytically described by a homogeneous fluid velocity $\mathbf{u}$, schematically shown in Fig.~\ref{figN}(a). We emphasize that UHSs are different from spin waves in which the former are nonlinear, spatially textured magnetization states while the latter are small-amplitude, linear perturbations of a magnetization state. Furthermore, UHSs are topologically protected by the in-plane magnetization's phase winding and concomitant large cone angle. This topological protection also gives rise to peculiar effects such as broken Galilean invariance~\cite{Iacocca2017} and the shedding of topology-conserving vortex-antivortex pairs from an impenetrable obstacle~\cite{Iacocca2017b}.

For the case of a finite-sized magnetic material, a canonical theoretical model is an effective one-dimensional, planar ferromagnetic channel subject to a non-equilibrium spin accumulation, or spin injection, at one edge. A solution to this model is a large-amplitude magnetization state exhibiting a spatially homogeneous precessional frequency and algebraic decay of fluid velocity as a result of damping~\cite{Konig2001,Sonin2010,Chen2014,Takei2014}, schematically depicted in Fig.~\ref{figN}(b). This solution is sometimes called a spin superfluid, a term originally proposed by Sonin~\cite{Sonin2010}, who was motivated by the fact that the order parameter for an easy plane ferromagnet is topologically identical to that for a superfluid. Such a magnetization state is similar to a UHS as it describes a large-amplitude, textured magnetization and a homogeneous precessional frequency. However, a notable difference is that the fluid velocity or, equivalently, the exchange flow is dissipated by damping along the channel. We refer to this state as a dissipative exchange flow, whereby a textured magnetization state exhibiting a well-defined precessional frequency is sustained by the balance between spin injection (forcing) and damping (dissipation). The use of the terminology dissipative exchange flow is motivated by other steady state excitations in magnetic materials, such as propagating and localized modes~\cite{Slonczewski1999,Slavin2005,Dumas2013,Iacocca2015} and dissipative droplets~\cite{Hoefer2010,Mohseni2013,Iacocca2014,Macia2014,Lendinez2015,Chung2016}, that are established by a local balance between forcing and dissipation. In contrast, the salient feature of dissipative exchange flows is that the balance is nonlocal; i.e., spin injection is established at the edge while dissipation occurs along the entire length of the channel. The main implication of the previous statement is that dissipative exchange flows can, in principle, be established in an arbitrarily long channel at the expense of the magnitude of the spatially homogeneous precessional frequency~\cite{Sonin2010,Takei2014}. It is important to emphasize that the precessional frequency can pump dc spin current into a suitable reservoir at the unforced edge of the channel, or any other location along the channel, and with equal efficiency everywhere. This defining property of the dissipative exchange flow constitutes a novel feature that may be useful for spintronic applications.

The theoretical studies on dissipative exchange flows to date have made a crucial assumption: axial symmetry. This assumption breaks down, for example, in realistic materials whose crystal structure establishes a magnetocrystalline anisotropy or in ferromagnetic channels (nanowires) whose shape will induce an effective in-plane anisotropy. From an energetic perspective, domain walls are favored by in-plane anisotropy~\cite{Kim2015}, in which case a train of domain walls with identical chirality or a soliton lattice will ensue~\cite{Sonin2010}, which can be interpreted as a symmetry-broken UHS. However, the excitation of a dissipative exchange flow upon spin injection in materials with in-plane anisotropy remains an open question. Within the linear, weak anisotropy regime, it has been speculated that symmetry-breaking terms are detrimental to dissipative exchange flows and would establish a minimum or threshold spin current density for their excitation~\cite{Konig2001,Chen2014,Takei2014}. Here, we provide a quantitative description of the onset and characteristic features of symmetry-broken dissipative exchange flows in ferromagnetic channels with in-plane anisotropy.

In this paper, we demonstrate the nature of hydrodynamic states in ferromagnetic materials with in-plane anisotropy. In the particular case of a ferromagnetic channel subject to spin injection, two characteristic features emerge. First, a critical spin injection threshold must be overcome to excite dissipative exchange flows, which we quantify in terms of the channel length and magnetic material parameters. Second, dissipative exchange flows exhibit hydrodynamic oscillations, described by a damped sine-Gordon equation. This implies that the precessional frequency develops harmonic overtones that reduce the efficiency of dc spin current pumped into an adjacent spin reservoir relative to a planar ferromagnetic channel. The dissipative exchange flow features mentioned above are also observed in the presence of nonlocal dipole fields by micromagnetic simulations. Moreover, we show that the spin injection threshold for a dissipative exchange flow can be exceeded by spin transfer torque~\cite{Slonczewski1996} from a finite-sized contact region, taking advantage of the contact-to-nanowire area ratio. These results establish design parameters and constraints that must be taken into account to pursue an experimental demonstration of dissipative exchange flows in ferromagnetic materials.

%-------------------------------
%-------------------------------

The paper is organized as follows: In Sec. II, we derive the dispersive hydrodynamic formulation for a symmetry-broken ferromagnet and show the relevant scalings to reduce the model to a damped sine-Gordon equation. In Sec. III, the existence of hydrodynamic states is explored for an unforced, extended thin film using periodic traveling wave solutions of the undamped sine-Gordon equation. The particular case of a channel subject to injection is studied in Sec. IV both analytically and numerically. In Sec. V we perform micromagnetic simulations as a proof of concept. Finally, we provide a discussion on the hydrodynamic interpretation of the phenomena and concluding remarks in Sec. VI.

\section{Analytical model}

Magnetization dynamics in ferromagnetic materials can be described by the Landau-Lifshitz equation
\begin{equation}
\label{eq:1}
  \frac{\partial\mathbf{m}}{\partial t} = -\mathbf{m}\times\mathbf{h}_\mathrm{eff} - \alpha\mathbf{m}\times\mathbf{m}\times\mathbf{h}_\mathrm{eff},
\end{equation}
expressed in dimensionless form where $\mathbf{m}=(m_x,m_y,m_z)$ is the normalized magnetization vector and $\alpha$ is the damping coefficient equivalent to the Gilbert damping parameter when $\alpha\ll1$. Time is scaled by $\gamma\mu_0M_s$, where $\gamma$ is the gyromagnetic ratio, $\mu_0$ the vacuum permeability, and $M_s$ the saturation magnetization; space is scaled by the exchange length $\lambda_{ex}$; and field is scaled by $M_s$. For the purposes of our work in all but the micromagnetic section, Sec. V, the normalized effective field $\mathbf{h}_\mathrm{eff}$ incorporates exchange, local (zero-thickness limit) dipole, and in-plane anisotropy along an arbitrary in-plane direction $\hat{\mathbf{k}}=(k_x,k_y)$
\begin{equation}
\label{eq:2}
  \mathbf{h}_\mathrm{eff} = \underbrace{\Delta\mathbf{m}}_\text{exchange} - \underbrace{m_z\hat{\mathbf{z}}}_\text{local dipole} + \underbrace{h_\mathrm{an}\left(k_xm_x\hat{\mathbf{x}}+k_ym_y\hat{\mathbf{y}}\right)}_\text{in-plane anisotropy}.
\end{equation}

In order to capture the full nonlinearity and exchange dispersion of Eqs.~\eqref{eq:1} and \eqref{eq:2} in an analytically tractable representation, it is possible to map the magnetization vector to hydrodynamic variables. In particular, we implement the transformation $n=m_z$ and $\mathbf{u}=-\nabla\Phi=-\nabla\arctan{(m_y/m_x})$, where $n$ is the longitudinal spin density and $\mathbf{u}$ is the fluid velocity.The fluid velocity plays the role of the texture's wavevector, suggesting an intimate relationship to the exchange length that typically scales the dispersion relation of small-amplitude spin waves~\cite{Stancil2009}.

Introducing the hydrodynamic variables into Eqs.~\eqref{eq:1} and \eqref{eq:2}, we obtain the dispersive hydrodynamic (DH) formulation of magnetization dynamics~\cite{Iacocca2017} with the addition of in-plane anisotropy. Considering a one-dimensional channel elongated in the $\hat{\mathbf{x}}$ direction, such that the one-dimensional fluid velocity is $u=\mathbf{u}\cdot\hat{\mathbf{x}}=-\partial_x\Phi$, the resulting DH equations are
\begin{subequations}
\label{eq:3}
\begin{eqnarray}
  \label{eq:31}
    \partial_tn &=& \partial_x\left[(1-n^2)u\right] + \frac{h_\mathrm{an}}{2}(1-n^2)(k_x-k_y)\sin{2\Phi}\nonumber\\&&+\alpha(1-n^2)\partial_t\Phi,\\ 
  \label{eq:32}
    \partial_t\Phi &=& -\left(1-u^2\right)n+\frac{\partial_{xx}n}{1-n^2}+\frac{n (\partial_xn)^2}{(1-n^2)^2}\nonumber\\&&-h_\mathrm{an}n\left(k_x\cos^2{\Phi}+k_y\sin^2{\Phi}\right)+\frac{\alpha\left[\alpha(1-n^2)\partial_t\Phi-\partial_tn\right]}{1-n^2}.
\end{eqnarray}
\end{subequations}
We emphasize that these equations represent an exact transformation of Eqs.~\eqref{eq:1} and \eqref{eq:2}. The change in density is driven by the flux
\begin{equation}
\label{eq:scd}
  q_s = -(1-n^2)u,
\end{equation}
the first term in the right-hand side of Eq.~\eqref{eq:31}. This dimensionless flux is identical to the equilibrium spin current density that results from non-collinear magnetic moments ($u\neq0$)~\cite{Saitoh2012}.

If we consider a small, but non-zero in-plane anisotropy in Eqs.~\eqref{eq:31} and \eqref{eq:32}, $0\ll h_\mathrm{an}\ll1$, it is possible to introduce the slow time, long wavelength, and small density scalings $T = \sqrt{h_\mathrm{an}} t$, $X = \sqrt{h_\mathrm{an}} x$, and $N=n/\sqrt{h_\mathrm{an}}$ to approximate Eq.~\eqref{eq:32} to leading order with $N=\partial_T\Phi$ and Eq.~\eqref{eq:31} by the damped sine-Gordon equation
\begin{equation}
\label{eq:1007}
  \partial_{TT}\Phi - \partial_{XX}\Phi + \frac{\alpha}{\sqrt{h_\mathrm{an}}} \partial_{T}\Phi + \frac{k_x-k_y}{2} \sin 2 \Phi = 0.
\end{equation}

Interestingly, it is possible to quench the effect of anisotropy in this limit when the relative angle between the anisotropy and the fluid velocity is $45$ degrees i.e., $k_x=k_y$. More generally, we here consider the anisotropy to be finite and aligned with the fluid velocity, i.e., $k_x=1$ and $k_y=0$. Different anisotropy geometries simply lead to a rescaling of time, space, and damping. Other symmetry breaking terms such as a small in-plane field may be introduced in Eq.~\eqref{eq:2} and interpreted hydrodynamically in a fashion similar to what has been presented above.

\section{Hydrodynamic states in symmetry-broken ferromagnets}

\begin{figure*}[t]
\centering \includegraphics[trim={0in 1in 0in 1.5in}, clip, width=6.5in]{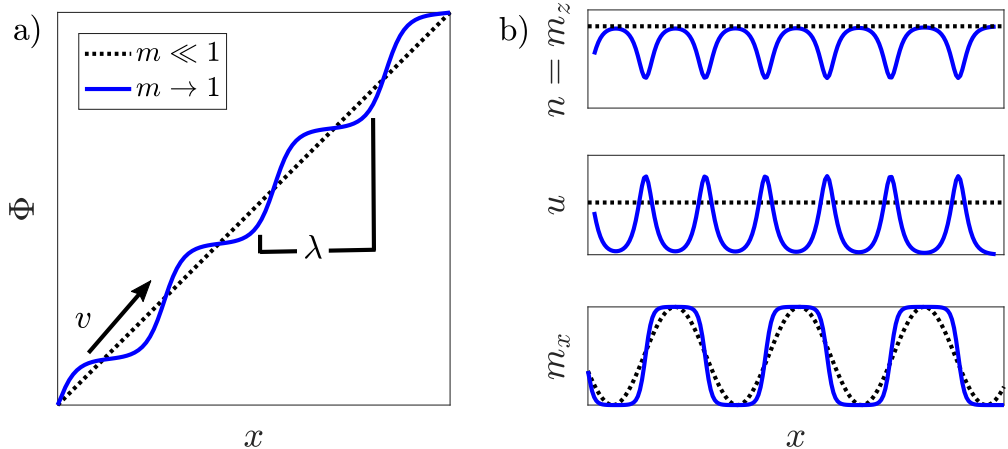}
\caption{ \label{fig1} (color online) (a) Schematic of oscillatory UHS phase solutions in the perturbed UHS ($m\ll1$) and soliton lattice ($m\rightarrow 1$) limits. (b) Corresponding fluid density $n$, fluid velocity $u$, and magnetization component $m_x$. }
\end{figure*}

We first study the existence of hydrodynamic-type solutions to Eqs.~\eqref{eq:31} and \eqref{eq:32} by analyzing the conservative limit, $\alpha=0$. In the case of axially symmetric, planar ferromagnets, both static, spin density waves (SDWs) and dynamic, uniform hydrodynamic states (UHSs) parametrized by a constant density and fluid velocity are supported~\cite{Iacocca2017}. The trigonometric terms arising from in-plane anisotropy in Eqs.~\eqref{eq:31} and \eqref{eq:32} modify the SDWs and UHSs. We can use Eq.~\eqref{eq:1007} and, e.g., Ref.~\onlinecite{Jones2013} to obtain approximate, traveling wave solutions in the coordinate $\xi =x-vt$, where $v$ is the velocity. The conservative ($\alpha=0$), dynamic solution of Eqs.~\eqref{eq:31} and \eqref{eq:32} for weak anisotropy ($0<h_\mathrm{an}\ll1$) can be approximately expressed as
\begin{equation}
\label{eq:3x}
  \cos{(2\Phi)} \sim \mathrm{sgn}(1-v^2)\left[-1 + 2\mathrm{sn}^2\left(\sqrt{\frac{h_\mathrm{an}}{m|1-v^2|}}(\xi-\xi_0),m\right)\right],
\end{equation}
and $n\sim v\sqrt{h_\mathrm{an}}\partial_x\Phi$, where sn is a Jacobi elliptic function, $\xi_0$ sets the initial phase, and $0<m<1$ is a parameter that determines the form of the solution. Equation~\eqref{eq:3x} represents a family of solutions, parametrized by $v$ and $m$, exhibiting spatially oscillatory density and fluid velocity with perturbed UHSs [$m\ll1$, $\mathrm{sn}^2(z,m)\sim\sin^2{(z)}$] and dynamic soliton lattices [$m\rightarrow1$, $\mathrm{sn}^2(z,m)\sim\tanh^2{(z)}$, repeated on the long wavelength proportional to $\ln{1/(1-m)}$] as limiting cases. These solutions are schematically depicted in Fig.~\ref{fig1}. Interestingly, it is possible to find period-averaged quantities for these oscillatory solutions leading to the long-wave dispersion relations
%-------------------------------
%-------------------------------
\begin{equation}
\label{eq:4x}
  v=-\frac{\bar{N}}{\bar{U}},\quad \bar{U}=\frac{-\mathrm{sgn}(\Delta\Phi)\pi}{\lambda},\quad\bar{\Omega}=-\bar{N}=v\bar{U},
\end{equation}
where $\Delta\Phi$ determines the precession orientation (positive is anti-clockwise), $\lambda=2K(m)\sqrt{m|1-v^2|/h_\mathrm{an}}$ is the oscillation wavelength, $K(m)$ is the complete elliptic integral of the first kind, $\bar{\Omega}$ is the frequency, and $\bar{N}$ and $\bar{U}$ are the mean density and fluid velocity, respectively. The nonlinear dispersion relation $\bar{\Omega}=-\bar{N}$ in Eq.~\eqref{eq:4x} agrees with that of the axially symmetric UHS in an averaged sense~\cite{Iacocca2017}. This identification implies that the symmetric UHS velocity, $v_\mathrm{UHS}$, is identical to the symmetry-broken UHS velocity, $v$. When $\bar{\Omega}=\bar{N}=v=0$, this static solution represents a symmetry-broken SDW whose symmetric counterpart was studied in Ref.~\onlinecite{Iacocca2017}. The above analysis demonstrates that hydrodynamic states exist in materials with in-plane anisotropy, featuring hydrodynamic oscillations that agree with axially symmetric UHSs and SDWs in an averaged sense.

\section{Spin injection through a symmetry-broken ferromagnetic channel}

We now consider a channel of length $L$ subject to spin injection polarized along the $\hat{\mathbf{z}}$ direction at the left edge of the channel. It is critical to find a hydrodynamic representation for spin injection. In general, this is achieved by adding a spin-transfer torque (STT) term to the right-hand side of Eq.~\eqref{eq:1} in the form~\cite{Slonczewski1996}
\begin{equation}
\label{eq:stt}
  \tau = -\mu\mathbf{m}\times\mathbf{m}\times\hat{\mathbf{z}},
\end{equation}
where $\mu$ is the dimensionless spin injection polarized along the $\hat{\mathbf{z}}$ component acting on the left edge of the channel, $x=0$. In hydrodynamic variables, Eq.~\eqref{eq:stt} only appears as a damping-like term $\mu(1-n^2)$ added to the right-hand side of Eq.~\eqref{eq:31}. This implies that spin injection results in a proportional equilibrium spin current density, Eq.~\eqref{eq:scd}, and that $\mu$ is proportional to the damping term $\alpha\partial_t\Phi$. In other words, it is possible to parametrize spin injection either in terms of an equilibrium spin current density, $q_s\neq0$ in Eq.~\eqref{eq:scd}, impinging on the channel's edge or a STT-induced magnetization precession, $\partial_t\Phi\neq 0$. In fact, both representations are physically equivalent from the perspective of non-equilibrium spin accumulation and resultant spin current flow across a magnetic / non-magnetic interface. In this section, we will parametrize spin injection hydrodynamically as an input flow $\bar{u}\propto\mu$ where we assume that the spin injection polarity induces a same-signed input flow. As we demonstrate below, the input flow establishes a dissipative exchange flow exhibiting a homogeneous precessional frequency. Note that the sign of the spin injection, or the direction of the input flow $\bar{u}$, merely dictates the clockwise or anti-clockwise precession of otherwise degenerate states. Further assuming free spin conditions and disregarding the effect of neighboring metallic layers from which spin currents can be injected $(x=0)$ or pumped $(x=L)$ for simplicity, we are left with the boundary conditions (BCs)
\begin{subequations}
\begin{eqnarray}
\label{eq:41}
  \partial_xn(0,t)=0, &\quad& \partial_xn(L,t)=0,\\
\label{eq:42}
	\partial_x\Phi(0,t)=-\bar{u}, &\quad& \partial_x\Phi(L,t)=0.
\end{eqnarray}
\end{subequations}

In the case of an isotropic planar ferromagnet, $h_\mathrm{an}=0$, and under appropriate long channel and weak injection approximations, Eqs.~\eqref{eq:31} and \eqref{eq:32} subject to Eqs.~\eqref{eq:41} and \eqref{eq:42} (equivalently Eq.~\eqref{eq:1007} with $k_x=k_y$) yield the approximate dissipative exchange flow~\cite{Sonin2010,Takei2014}
\begin{equation}
\label{eq:7}
  u_s = \bar{u}(1-\frac{x}{L}),\quad\Omega_{s}=\frac{\bar{u}}{\alpha L},\quad n_s=-\Omega_{s},
\end{equation}
where the subscript ``s'' indicates an axially symmetric solution with the fluid velocity $u_s$, fluid density $n_s$, and precessional frequency $\Omega_s$. Dissipative exchange flows exhibit a uniform precessional frequency for any nonzero input flow $\bar{u}$. Notably, the dispersion relation of a symmetry-broken UHS is maintained but the wave velocity is space dependent, $v_s(x)=-n_s/u_s(x)$. The linear decay of fluid velocity $u_s$ along the channel manifests as a spatial increase of the in-plane magnetization wavelength, see Fig.~\ref{figN}(b). It is important to emphasize that the balance between the edge input flow and dissipation along the channel that sustains dissipative exchange flows manifest in the precessional frequency as the ratio $\bar{u}/\alpha L$.

For the nonzero but small anisotropy regime, $0<h_\mathrm{an}\ll1$, we study the approximate damped sine-Gordon Eq.~\eqref{eq:1007} subject to Neumann boundary conditions
\begin{equation}
\label{eq:1008}
  \partial_X\Phi(0,t) = - \frac{\overline{u}}{\sqrt{h_\mathrm{an}}}, \quad
  \partial_X\Phi(\sqrt{h_\mathrm{an}}L,t) = 0,
\end{equation}
modeling the ferromagnetic channel in the low frequency, long wavelength regime.

%-------------------------------
\begin{figure*}[t]
\centering \includegraphics[trim={0in 0.3in 0in 0in}, clip, width=6.5in]{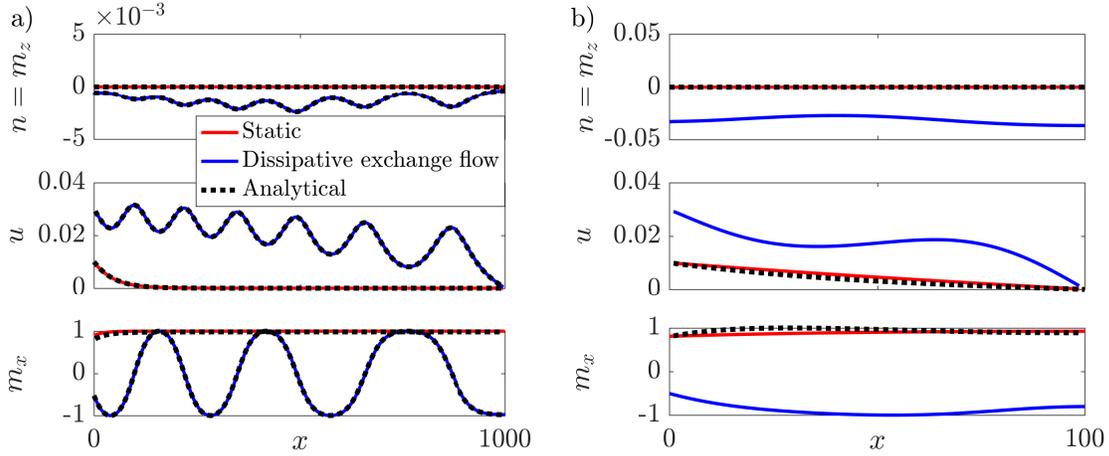}
\caption{ \label{figN2} (color online) Static, non-oscillatory states (solid red curves, $\bar{u}=0.01$) and oscillatory dissipative exchange flows (solid blue curves, $\bar{u}=0.03$) for $n$, $u$, and $m_x$ in ferromagnetic channels with in-plane anisotropy. The parameters are $\alpha=0.01$ and $h_\mathrm{an}=5\times10^{-4}$, from which $l_\mathrm{dw}=44.7$. For the long channel, (a) $L=1000$, the static solution decays exponentially and it is well described by Eq.~\eqref{eq:x1} (dashed black curves). The presented dissipative exchange flow is taken at a particular instant of time, and is accompanied by oscillations both in $n$ and $u$ that are well described by the sine-Gordon equation, shown by dashed black curves. For the short channel, (b) $L=100$, the static solution decays approximately linearly and is also well-described by Eq.~\eqref{eq:x1} (dashed black curves). The dissipative exchange flow is still oscillatory. }
\end{figure*}
%-------------------------------
Equation~\eqref{eq:1007} admits the trivial solution $\Phi=0$ and $N=0$ that reflects the static ground state along the easy axis imposed by the anisotropy when $\bar{u}=0$. For a finite input flow, $\bar{u}\neq0$, this ground state is perturbed. If the input flow is small enough, a static solution ($\partial_T\Phi=0$) can be obtained analytically by considering the generalized SDW solution, Eq.~\eqref{eq:4x} with $v=0$. The application of the boundary conditions \eqref{eq:1008} yield
\begin{subequations}
\begin{eqnarray}
 \label{eq:x1}
    \cos{(2\Phi)} &=& -1 + 2m\mathrm{sn}^2\left(X-\sqrt{h_\mathrm{an}}L+K(m),m\right),\\
\label{eq:x2}
  \frac{\bar{u}^2}{h_\mathrm{an}} &=& m\left[1-\mathrm{sn}^2\left(\sqrt{h_\mathrm{an}}L-K(m),m\right)\right].
\end{eqnarray}
\end{subequations}
Given $\bar{u}$, $h_\mathrm{an}$, and $L$, Eq.~\eqref{eq:x2} determines $0<m<1$ and Eq.~\eqref{eq:x1} yields the static spatial profile. From this solution and the spatial scaling $X=\sqrt{h_\mathrm{an}}x$, we identify the typical domain wall length scale, $l_\mathrm{dw}\propto h_\mathrm{an}^{-1/2}$. Numerical solutions for this regime can be found by solving Eq.~\eqref{eq:1} and \eqref{eq:2} subject to Eqs.~\eqref{eq:41} and \eqref{eq:42} in time, initialized at the ground state $\mathbf{m}=(1,0,0)$, until a static steady state is reached. We use $\alpha=0.01$, $h_\mathrm{an}=5\times10^{-4}$ and channel lengths $L=1000$ and $L=100$ for panels (a) and (b), respectively. The long-time solution for $n$, $u$, and $m_x$ obtained with the input flow $\bar{u}=0.01$ are shown by red solid curves in Fig.~\ref{figN2} and is in excellent agreement with the analytical solution Eq.~\eqref{eq:x1} (dashed black curves). For a long enough channel, $L\gg l_\mathrm{dw}$ [Fig.~\ref{figN2}(a)], the solution is exponentially decaying in phase and therefore fluid velocity $u=-\partial_x\Phi$ as well. When $L\approx l_\mathrm{dw}$ [Fig.~\ref{figN2}(b)], the static solution exhibits approximately linear decay. These decaying solutions are fundamentally different from exponentially decaying spin waves. Here, the solution is static so there is no energy dissipation associated with magnetization precession about the local equilibrium direction. In contrast, the injected energy tilts the magnetization near the left edge of the channel towards the hard axis. This implies that in-plane anisotropy acts as an energy barrier that prevents the introduction of topological phase winding and its associated hydrodynamic flow.

A dynamic dissipative exchange flow is recovered when the input flow is sufficient to overcome the static regime, shown by solid blue curves in Fig.~\ref{figN2} for $\bar{u}=0.03$. In contrast to the symmetric dissipative exchange flow, Eq.~\eqref{eq:7}, both the density $n$ and fluid velocity $u$ exhibit oscillations on the domain wall and input flow length scales, in agreement with oscillatory hydrodynamic states, Eq.~\eqref{eq:3x}. These oscillations are dominated by large-amplitude magnetization precession, as shown by $m_x$ in Fig.~\ref{figN2}(a), bottom panel. However, the entire solution coherently precesses at a fixed, fundamental frequency, exhibiting higher harmonic content due to nonlinearity. This implies that the oscillations will also manifest spectrally. The above simulations expose a competition between the different length scales that exist in the system, namely, the domain wall length scale, the channel length, and the dissipative exchange flow wavelength proportional to $l_\mathrm{dw}$, $L$, and $\bar{u}^{-1}$, respectively.  Animated versions of the dissipative exchange flows shown in Fig.~\ref{figN2} can be found in the supplemental videos 1 and 2, respectively.

\begin{figure}[t]
\centering \includegraphics[trim={0in 0.3in 0in 0.1in}, clip, width=3.5in]{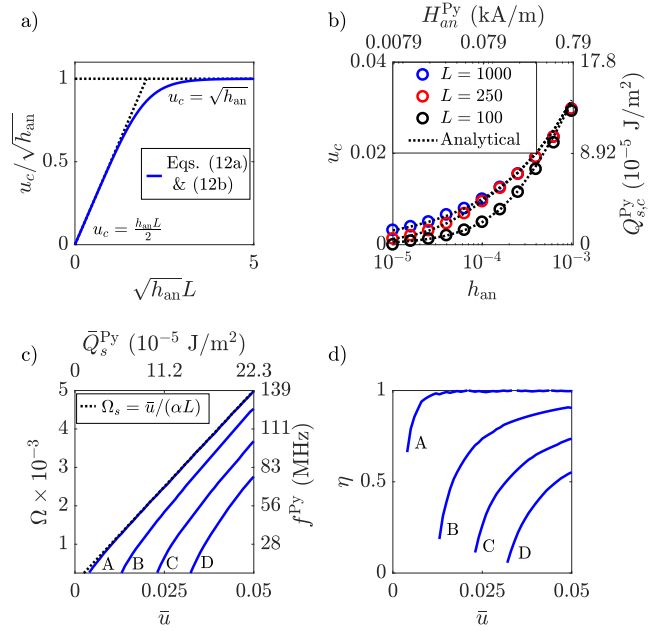}
\caption{ \label{fig2} (color online) (a) Threshold current as a function of in-plane anisotropy and channel length (solid blue curve) as well as the short and long channel asymptotic limits (dashed black lines). (b) Input flow threshold for dissipative exchange flows as a function of in-plane anisotropy magnitude for different choices of $L$ and $\alpha=0.01$, computed from Eq.~\eqref{eq:x2} (dashed black curves) and by numerically estimating the onset of hydrodynamic solutions from an initial value problem for Eqs.~\eqref{eq:31} and \eqref{eq:32} (circles). The right and top axes show, respectively, the threshold spin current density and anisotropy field in physical units utilizing magnetic parameters of Py. (c) Numerically determined frequency, $\Omega$, as a function of $\bar{u}$ for different in-plane anisotropy magnitudes labeled $A$: $h_\mathrm{an}=10^{-5}$, $B$: $h_\mathrm{an}=1.6\times10^{-4}$, $C$: $h_\mathrm{an}=5\times10^{-4}$, and $D$: $h_\mathrm{an}=10^{-3}$ in a channel of length $L=1000$. The slope of the linear dependence for the axially symmetric ferromagnet, Eq.~\eqref{eq:7}, is shown by a dashed black line. The right and top axes show, respectively, the equivalent frequency and spin current density in physical units utilizing magnetic parameters of Py. The corresponding efficiency calculated using Eq.~\eqref{eq:11} is shown in (d). }
\end{figure}

The threshold (or critical) input flow, $u_c$, can be determined from the existence of the static solution, Eqs.~\eqref{eq:x1} and \eqref{eq:x2}. In general, the critical input flow can be found numerically by solving  the transcendental equation~\eqref{eq:x2}, which identifies the maximum allowed $\bar{u}$ for a given channel length $L$. This is shown by the solid blue curve in Fig.~\ref{fig2}(a). The critical input flow exhibits two asymptotic limits
\begin{equation}
\label{eq:10}
  u_c = \begin{cases} \frac{h_\mathrm{an}L}{2}\quad,& L\ll l_\mathrm{dw}\\ \sqrt{h_\mathrm{an}}\quad,&L\gg l_\mathrm{dw}\end{cases},\quad Q_{s,c} = -\mu_0M_s^2\lambda_{ex}u_c
\end{equation}
where we show both the dimensionless critical input flow $u_c$ and its conversion to a dimensional equilibrium spin current density $Q_{s,c}$ in SI units, J/m$^2$, under the assumption that $n$ is small. These limiting behaviors are shown in Fig.~\ref{fig2}(a) by dashed black lines. Physically, the critical input flow is the injection necessary to tilt the magnetization to the hard axis, $\Phi=\pm\pi/2$, at which point the magnetization can continuously precess and establish a dynamic solution.
%-------------------------------
%-------------------------------
For the parameters of Fig.~\ref{figN2}a and \ref{figN2}b, $u_c=0.0224$ and $u_c=0.0204$, respectively. Numerically, it is possible to directly estimate $u_c$ as a function of $h_\mathrm{an}$ by solving the fully nonlinear Eqs.~\eqref{eq:1} and \eqref{eq:2} subject to Eqs.~\eqref{eq:41} and \eqref{eq:42} initialized at the ground state $\mathbf{m}=(1,0,0)$, and seeking the transition between a static and a dynamic regime. The numerical estimates are shown in Fig.~\ref{fig2}(b) by circles representing different choices for $L$ while maintaining $\alpha=0.01$. The computed threshold is shown by dashed black curves and quantitatively agrees with the numerical results. To gain intuition on the physical spin current densities required to excite a dissipative exchange flow, the right and top axes in Fig.~\ref{fig2}(b) are shown in physical units considering typical Permalloy (Py, Ni$_{80}$Fe$_{20}$) material parameters: $M_s=790$~kA/m and exchange stiffness $A=10$~pJ/m. For further reference, the largest critical spin current density for a Py anisotropy of $H_{an}=400$~A/m is $Q_{s,c}^\mathrm{Py}\approx9\times10^{-5}$~J/m$^2$. In comparison, the largest critical spin current density for a $29$ nm thick YIG film with $M_s=130$~kA/m, $H_{an}=1.9$~kA/m~\cite{Wang2014b}, and exchange stiffness $A=5$~pJ/m is comparable at $Q_{s,c}^\mathrm{YIG}\approx5.5\times10^{-5}$~J/m$^2$. It is important to recall that these estimates indicate the amount of angular momentum necessary to tilt the in-plane magnetization towards the hard axis. These high spin current density thresholds can be partially mitigated by working with shorter channels, as shown in Fig.~\ref{fig2}(a). Alternatively, utilizing a finite-sized region placed on top of the channel to induce STT makes it possible to effectively achieve such high spin current densities by inducing magnetization precession with experimentally moderate charge current densities.

Above threshold, a symmetry-broken oscillatory dissipative exchange flow is established along the channel. Intuitively, the oscillations require energy to be sustained, subtracting from the total energy pumped by the input flow $\bar{u}$. This energy distribution manifests as harmonic overtones of the precessional frequency $\Omega$ that redshift the fundamental frequency as a function of $h_\mathrm{an}$. By solving Eqs.~\eqref{eq:31} and \eqref{eq:32} numerically, it is possible to estimate the fundamental precessional frequencies as a function of the input flow $\bar{u}$, shown by solid blue curves in Fig.~\ref{fig2}(c) for several in-plane anisotropy magnitudes. The right and top axes are shown in physical units for Py, where $2\pi f=\gamma\mu_0M_s\Omega$. Note that the dimensional frequencies are in the MHz range. This corresponds to the fact that the magnetization in a dissipative exchange flow mostly lies in the plane, leading to a small local dipole field contribution to drive the precession in the absence of an external field. Above threshold, the frequency increases in a nonlinear fashion and asymptotically approaches the slope of the linear, axially symmetric solution $\Omega_{s}=\bar{u}/(\alpha L)$, shown as a dashed black line in Fig.~\ref{fig2}(c). Recalling that magnetization precession can pump spin current to an adjacent metallic reservoir, it is possible to define a spin pumping efficiency, $\eta$, as
\begin{equation}
\label{eq:11}
  \eta = \frac{\Omega}{\Omega_{s}}=\frac{Q_{s,p}}{\bar{Q}_s},
\end{equation}
where $\bar{Q}_s$ is the input spin current density and $Q_{s,p}$ is pumped spin current density that could be determined by inverse spin Hall measurements~\cite{Hoffmann2013} from a neighboring heavy metal, in which case different boundary conditions must be considered depending on the location of the spin reservoir. The efficiencies for the curves in Fig.~\ref{fig2}(c) are shown in Fig.~\ref{fig2}(d). Note that the efficiency can approach unity. This is because we define the efficiency relative to the axially symmetric solution that already takes into account the balance between spin injection and damping in establishing the steady state magnetization precession.

\section{Micromagnetic simulations in a symmetry-broken ferromagnet}

%-------------------------------
\begin{figure}[t]
\centering \includegraphics[trim={0in 0.in 0in 0.in}, clip, width=3.5in]{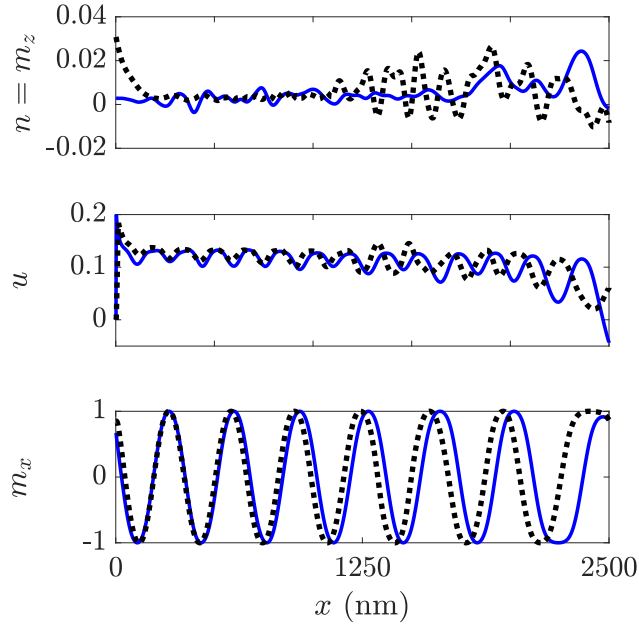}
\caption{ \label{figN3} (color online) Dissipative exchange flows obtained micromagnetically for a Py nanowire where magnetization precession close to the left edge is sustained by spin transfer torque from an area with dimensions $1.2$~nm~$\times~100$~nm subject to a charge current density $\bar{J}_c=2\times10^{10}$~A/m$^2$ (solid blue curves) and dimensions $12$~nm$~\times~100$~nm subject to a charge current density $\bar{J}_c=2\times10^{9}$~A/m$^2$ (dashed black curves). The curves are snapshots of the center-line along the nanowire obtained at times such that the fluid velocities partly overlap close to the left edge. }
\end{figure}
%-------------------------------

The above analytical results can be validated by micromagnetic simulations utilizing a local dipole field. However, we note that in-plane anisotropy can arise from the shape of an elongated channel by considering nonlocal dipole fields that are not incorporated in the analytical framework studied above. As a proof of concept for the validity of our local dipole field results, we run micromagnetic simulations in MuMax3~\cite{Vansteenkiste2014} for a Py channel of dimensions $2500$~nm~$\times~100$~nm~$\times~1$~nm. Spin injection is modeled as a symmetric STT~\cite{Slonczewski1996} impinging on a $1.2$~nm~$\times~100$~nm area located on the top left edge of the channel~\cite{Chen2014}. As mentioned above, STT induces magnetization precession and, therefore, parametrizes the spin injection $\mu$ in Eq.~\eqref{eq:stt}. In order to micromagnetically model the charge to spin current density transduction, we perform simulations with local dipole field (axially symmetric ferromagnet) and fit the STT polarization $P$ to match the axially symmetric frequency $\Omega_{s}$ as a function of a charge current density $J_c$. We obtain $P=0.65$. The incorporation of shape anisotropy leads to a total in-plane anisotropy of $948$~A/m for a channel of length $2500$~nm (see, e.g., Ref.~\onlinecite{Getzlaff2008}). The corresponding dimensionless parameters are $L=500$ and $h_\mathrm{an}=1.2\times10^{-3}$, which leads to $u_c=0.035$ or ${J_c\approx5.7\times10^{9}}$~A/m$^2$ in physical units. Imparting a charge current density of ${\bar{J}_c=2\times10^{10}}$~A/m$^2$ (equivalently $\bar{u}=0.1976$), an oscillatory dissipative exchange flow with fundamental frequency $178.52$~MHz ($\eta=0.16$) is excited, as shown by the solid blue curve snapshots in Fig.~\ref{figN3}. See supplementary video 3 and 4 for an animated version.

The charge to spin current density transduction is also enhanced by a geometric factor $A=w/t$, where $w$ is the width of the STT area and $t$ the thickness of the channel. For the dimensions shown above, $A=1.2$. We have micromagnetically verified that a STT region of $12$~nm in width yields an order of magnitude reduction in the required threshold charge current density relative to the $1.2$~nm wide STT region while maintaining the same efficiency and qualitative features. A snapshot of the resulting dissipative exchange flow with fundamental frequency $193$~MHz is shown by dashed black curves in Fig.~\ref{figN3}. The relatively high charge current densities with respect to the threshold charge current density to induce dissipative exchange flows suggest a further shift of the threshold when nonlocal dipole fields are included and can strongly influence the magnetization near the edges of the channel. In fact, the supplementary videos 3 and 4 show evidence that the nanowire's left and right edges nucleate and annihilate solitonic features, respectively. Despite such additional dynamics, these simulations demonstrate that broken axial symmetry is not a fundamental constraint for the existence of dissipative exchange flows.

\section{Discussion and conclusion}

We have shown that dissipative exchange flows exist in ferromagnetic channels with in-plane anisotropy, i.e., broken axial symmetry. We quantitatively determined the injection threshold as a function of material parameters, corresponding to that necessary for a spin-current-driven tilt of the magnetization along the hard axis. For spin injection above threshold, oscillatory solutions are obtained, whereby the magnetization temporal precession exhibits higher harmonic content that reduces the spin pumping efficiency when compared to the axially symmetric case.

The dispersive hydrodynamic formulation allows us to draw an analogy for the dissipative exchange flows described above with hydrodynamics. Spin injection can be viewed as fluid flow injected from a nozzle into a pipe. In-plane anisotropy acts in two different ways: first, as a lift-check valve at the exit of the nozzle, establishing a velocity (or pressure) barrier, and second, as a periodic corrugation in the pipe that leads to an oscillatory fluid density and velocity. However, this analogy is limited as a fluid interpretation disregards the peculiarities of magnetization dynamics. Notably, the flow experiences constant deceleration to damping while maintaining the density independently of in-plane anisotropy; see Fig.~\ref{fig1}(c). Interestingly, the hydrodynamic interpretation of magnetization dynamics here is described by the phase $\Phi$ or exchange flow velocity potential ($\mathbf{u}=-\nabla\Phi$), which admits the nonzero precessional frequency $\Omega=\partial_t\Phi$ as an observable that is determined by the magnetic analog of Bernoulli's equation~\cite{Iacocca2017}. This is in contrast to classical fluids where the velocity potential is obtained from the fluid velocity under the premise of irrotational flow and is not a physical observable.

Finally, micromagnetic simulations qualitatively agree with the analytical results in the presence of nonlocal dipole fields which will inevitably exist at the channel's edges. We further showed numerically that spin transfer torque from a finite-sized region placed on top of the channel can sustain dissipative exchange flows at experimentally accessible charge current densities.

\begin{acknowledgments}
E.I. acknowledges support from the Swedish Research Council, Reg. No. 637-2014-6863. M.A.H. is partially supported by NSF CAREER DMS-1255422.
\end{acknowledgments}

\end{document}